# How much do different ways of calculating percentiles influence the derived performance indicators? – A case study

**Michael Schreiber**

*Institute of Physics, Chemnitz University of Technology, 09107 Chemnitz, Germany*

*E-mail: schreiber@physik.tu-chemnitz.de , Tel.: +49-37153121910, Fax: +49-37153121919*

Bibliometric indicators can be determined by comparing specific citation records with the percentiles of a reference set. However, there exists an ambiguity in the computation of percentiles because usually a significant number of papers with the same citation count are found at the border between percentile rank classes. The present case study of the citations to the journal Europhysics Letters (EPL) in comparison with all physics papers from the Web of Science shows the deviations which occur due to the different ways of treating the tied papers in the evaluation of the percentage of highly cited publications. A strong bias can occur, if the papers tied at the threshold number of citations are all considered as highly cited or all considered as not highly cited.

**Introduction**

Bibliometric indicators for the evaluation of performance in terms of citations have been customarily based on average values. Such approaches suffer from the usually highly skewed citation distribution, dominated by highly-cited papers so that the arithmetic mean strongly deviates from the median. As an alternative to mean-based indicators the idea of looking at the position of a publication within a reference distribution has recently lead to an intensive discussion of percentile-based indicators (Bornmann, Leydesdorff, & Mutz, 2012; and references therein). Such indicators are based on percentile rank classes (PRCs) which comprise all publications with certain citation frequencies. The borders between the PRCs correspond to a threshold number of citations in the reference set. For example, for the highly-cited-publications (HCPs) indicator which refers to the top 10% most frequently cited papers the threshold is determined so that 90% of all papers in the reference set have received less (or: less or equal) citations. Here a problem with the determination of percentiles becomes clear: it is not obvious whether the paper in question is included in the paper count or not, hence the choice between *less* citations and *less or equal* citations in the comparison. If many of papers at the threshold received the same number of citations, large differences in the number of papers which are attributed to the PRCs can occur (Schreiber, 2012b). On the other hand, small changes in the reference set can lead to inconsistent behavior of the determined indicators (Schreiber, 2012a).

The recently introduced fractional scoring method (Schreiber, 2012b) treats the problem of the tied papers by fractionally assigning them to more than one percentile value and thus to more than one PRC. In this way one always obtains exactly 10% HCPs. I have shown (Schreiber, 2012b) strong deviations in the *R*6 indicator for a small model reference set due to the different counting rules. In the present investigation I extend that study to a large reference set of real (i.e., not model) data taken from the Web of Science. It is the purpose of this work to show how strongly the different possibilities of determining percentiles and the ambiguous ways of assigning tied papers to the PRCs influence the HCPs indicator. Specifically I utilize the results to describe the performance of the journal Europhysics Letters (EPL) as an example case in order to demonstrate how relevant these deviations may be for performance evaluations. The choice of this example is motivated by my personal interest in this journal, because as Editor-in-Chief of EPL I want to evaluate the performance of EPL. As the evaluation results of the present investigation for the example set are rather close to the reference set, I believe that the journal is typical for a multidisciplinary journal in physics and that concerning the percentile-based evaluation procedure it can be considered to be representative also for more specific subfields as well as for other fields, with the exception of extremely highly cited and very lowly cited journals.

Of course, there are many other ways of evaluating the performance of a journal or a researcher in terms of the publication record. Probably the simplest one is just counting the number of publications. Counting the number of citations is more related to the impact of the publications. Determining the ratio of the number of citations and the number of publications leads to the well-known impact factor of a journal. In all these cases a bias towards or against specific (sub)fields is likely. Comparing the citation record with a reference set makes it easier to treat different fields of science in a completely equal way, because their specific citation distributions are taken into account. A straightforward way of such a comparison with a reference set is to analyze the citation records in terms of percentiles or quantiles and to attribute the publications of the reference set as well as the publications in the example set which is to be evaluated to PRCs. Then the share of publications in these PRCs can be compared between the reference set and the example set. Using different weights for the different PRCs the deviation of the performance of the example set from the reference set can be quantified.

The present investigation contributes to the recent discussion about the best procedure for this comparison (Bornmann, Leydesdorf & Mutz, 2012; Waltman & Schreiber, 2012; and references therein). Percentiles and PRCs have become a standard instrument in bibliometrics and it is not the purpose of the present paper to argue for or against their use in contrast to some other approach from the arsenal of possible statistics. Nor is it the purpose of the present manuscript to argue in favor of or against reducing the evaluation of the performance of a researcher or a journal to a single number derived from a single indicator. In this respect I am rather skeptical about any such procedure, but one has to take it as a fact that such evaluations have become very popular not only among administrators, but also among scientists especially on evaluation and recruiting committees, as well as among bibliometricans and scientometricans. Therefore it appears necessary to me to have a critical look at such indicators and to try to identify weaknesses and to try to improve the determination of such indicators.

**The reference set and its percentile-based evalution**

The citation distributions for the physics papers have been harvested from the Web of Science for 25 publication years (1986-2010) comprising 2,723,019 papers with 43,878,496 citations until the end of 2011. For each year the papers are sorted according to the number of citations.

I restrict the subsequent investigation to the HCPs indicator. For its computation one considers two PRCs, namely the bottom 90% and the top 10%, generically attributing weights 0 and 1 to them, respectively. This means that in each sorted list the 10% most cited papers, that is those in the top deciles get a weight 1, the others none. Thus the indicator reflects the share of HCPs. In order to avoid a misunderstanding, it should be noted that in the following I calculate percentiles starting with zero citations so that high percentile values correspond to high citation numbers. Accordingly, the 90% threshold determines the 10% most cited papers.

In Figure 1 the values of the citation frequency at the 90% threshold are given and the resulting percentages of papers below, at, and above the threshold are visualized. As an example, let us consider the publication year 1986: there are 60,468 out of 67,206 papers with less than 54 citations in the reference set, which means 89.97%. On the other hand 6579 papers, i.e. 9.79% have received more than 54 citations. Thus 54 citations mark the 90% threshold as indicated in Fig. 1. Exactly at the threshold one finds 159 papers (0.24%) Similarly, for the publication year 2010 the threshold is given by 7 citations, because 130,801 of 147,894 papers (i.e. 88.44%) have obtained less than 7 citations and 9.23%, namely 13,703 papers were cited 8 or more times. 3390 papers (2,29%) were cited exactly 7 times. Altogether at the 25 thresholds 0.56% of all papers are tied. This leads to the question, what the exact limit of the top decile is. Different counting rules lead to different results, some of which will be discussed below. Leydesdorff and Bornmann (2011) proposed the "counting rule that the number of items with lower citation rates than the item under study determines the percentile". Ranking the publications in increasing order by their number of citations and then attributing a percentile ($r-1)/N*100$ where $r$ is the rank corresponds to this lower-citation-frequency counting rule labelled "L" below. All "tied, that is, precisely equal, numbers of citations thus are not counted as fewer than" (Leydesdorff, Bornmann, Mutz, and Opthoff, 2011). This means that the factual threshold is always above 90%. For example consider the year 2010: The L rule means that the above mentioned 3390 papers at the threshold are assigned to the percentile value 88.44 so that 134,191 papers (90.73%) fall below the border, what makes 90.73% the factual threshold.

If one includes the item under study into the number of items to compare with (Rousseau, 2012) the percentile for a single publication is given by $r/N*100$ differing only by $1/N$ (= $100/N$ %) so that the effect is negligible for datasets as large as in the present study. However, tied papers are all assigned to the largest percentile value, i.e. in the example for the publication year 2010 the 3390 tied papers with exactly 7 citations are all given the percentile value 90.73 and thus attributed to the top decile. Therefore now the factual threshold is 88.44%. Obviously in this way indicated "E" (for equal citation frequency) below, one always gets a factual threshold below 90%.

The two rules L and E can be considered as the extreme possibilities, defining an uncertainty interval for the factual threshold as visualized in Figure 1. A compromise denoted "C" below, is to assign the average percentile value to publications with equal citation counts (Bornmann et al., 2012) or to average the two factual citation thresholds (Leydesdorff, 2012) what in the present example means 89.59%. This value is then utilized to decide whether the papers at the threshold are considered to be below or above the border. The factual threshold therefore is that of the two above extremes which is closest to 90%, in the present example 90.73%, because the average value of 89.59% puts all the tied papers below the 90% border line as for the L rule above.

The uncertainty of 2.29 percentage points between the two extreme rules L and E is relevant in so far as it reflects the uncertainty whether the 3390 papers belong to the top decile or not and whether they are therefore counted as highly cited or not. Of course, publications in recent years had less time to be cited. Correspondingly the citation frequencies at the 90% threshold are lower in recent years as indicated in Fig. 1, as usual the number of tied papers is the larger, the smaller the citation frequency is, and thus the proportion of tied papers is larger (see Fig. 1). Therefore it is to be expected that the above mentioned uncertainty whether the tied papers are considered as highly cited or not is most severe for the more recent publication years. Nevertheless, even for the publication year 1986 with a threshold of 54 citations, the 159 tied papers at this threshold lead a deviation of 0.21 percentage points from 90% for the L counting rule and -0.03 percentage points for E and C. On average, for the L, E, and C method the deviations from 90% amount to 0.22, -0.27, 0.11 percentage points, respectively.

In the fractional scoring approach (Schreiber, 2012b; Waltman & Schreiber, 2012) called "F" below, the publications at the threshold are fractionally assigned so that the exact shares of papers below and above the border are achieved. This means that the 2.29% border line papers in the example are divided into 90% - 88.44% = 1.56% below and 90.73% - 90% = 0.73% above the threshold. Thus 2303.6 papers (namely 1.56% / 2.29% of the 3390 tied papers) are considered to fall below and 1086.4 (= 0.73% / 2.29% of 3390) papers are considered to lie above the threshold. These 1086.4 papers are therefore assigned to the top decile, which thus comprises 13,703 + 1086.4 = 14789.4 papers, that is exactly 10% of the total number of papers from the year 2010. An alternative interpretation is that **all** the 3390 papers at the threshold are fractionally counted so that a share of 1.56% / 2.29% = 0.68 of **each** of the tied papers falls below and a share of 0.73% / 2.29% = 0.32 lies above the border. This means that a fraction of 0.32 of each of the tied papers belongs to the top decile; of course all papers with higher citation counts belong completey to the top decile. A different, though equivalent view is that the uncertainty *interval* visualized in Fig. 1 is assigned to the tied papers and that this interval is divided into the two parts below and above the 90% border line, so that a fraction of 0.32 falls above the border. In this case the alternative interpretation is that the interval is attributed with a weight of 0.32 to the top decile.

**The datasets for the case study and their performance**

The citation distributions for the publications in the journal Europhysics Letters (EPL) have also been obtained from the Web of Science. In 25 years there are 12774 papers with 217,929 citations.

In Figure 2 the shares of papers below, at, and above the thresholds are visualized. These thresholds are defined by the reference datasets but their positions are determined by the papers with corresponding citation frequencies in the example datasets. At the 25 thresholds 0.63% of all papers are tied.

The results of the HCPs indicator can be directly read from Figure 2. For example, 91.38% of the papers from 2010 have been cited less than 7 times (i.e. the 90% threshold given by the reference set). 1.34% got exactly 7 citations, and 7.28% received 8 or more citations. The upper boundary of the uncertainty interval which is defined by the border line papers corresponds to the L rule counting the percentage above this boundary (100% - 92.72% = 7.28% HCPs in 2010). Likewise, the lower boundary of the uncertainty interval reflects the result of the E counting rule (e.g. 100% - 91.38% = 8.62% HCPs in 2010). For the fractional scoring the uncertainty interval has to be divided into the same fractions as for the reference set, as indicated in Figure 2. This means that a share of 0.32 of the 1.34% border line papers is counted as highly cited. For the C rule, the result is equal to the L or E value according to the decision that was made in the evaluation of the reference set. Effectively this means that the result for the C rule is equal to the L or E value, depending on which of these is closer to the F value. For all counting rules strong fluctuations occur showing that the performance changes substantially from year to year.

In most cases the change from year to year which is represented by the position of the uncertainty interval is much larger than its width which is due to the different counting rules. For the recent years these differences become larger because publications had less time to be cited, which means a lower threshold citation frequency corresponding to a larger number of tied papers. Comparing with the fractional scoring, because this yields (by construction) exactly 10% HCPs for the reference set, the deviations from the F results are presented in Figure 3. Whether one considers the differences of several percent as relevant or not is an open question. In my view, they are unpleasantly large and should be avoided. I consider these differences as relevant, when for example in the case of the year 2009 either 7.86 % or 9.51 % of all EPL papers turn out to be highly cited. I think this is a substantial difference.

Another problem is that small modifications in a reference set can lead to relatively large changes of the indicators. As an example consider the year 1994. There are 97382 papers in the reference dataset, the threshold number of citations is 48, with 87640 papers having less than 48 citations while 268 papers received 48 citations each. Consequently, the uncertainty of the indicator (that is the difference between the results of the L and E counting rules) amounts to 0.27 percentage points, reaching from 9.73% to 10.004%, see Table 1. If 4 of the papers at the threshold were cited once less, then the threshold decreases to 47 citations with 87324 papers below and 320 at the new threshold. Now the different counting rules lead to an uncertainty between 9.998% and 10.33%, or 0.33 percentage points. As a consequence, the L and E values of the HCPs indicator for Europhysics Letters in 1994 rise substantially, compare Table 1. It is certainly a strange behavior that a minor modification (4 of 1,983,982 citations) in the reference set can lead to such a large change of these indicators for the example dataset. The C result remains unchanged. The F result increases slightly what is reasonable because it reflects the small deterioration in the reference set. The strange effect of the described small modification in the reference set shows that it is relevant which counting rule has been chosen for the analysis.

A similar behavior can be observed whenever one of the extreme indicator values, i.e. a boundary of the uncertainty interval, in Figure 1 comes close to 90%. As another example, I mention the year 2003 where 463 papers with 38 citations are tied at the 90% threshold. If only two of these had received one less citation, the threshold would be 37 citations and the uncertainty interval would change due to this modification as denoted in Table 2. For the example dataset the range of 5.45% – 6.03% determined for the original situation would be replaced after the modification by a single value of 6.03%, because no Europhysics Letters paper from 2003 received exactly 37 citations. In this case the L values indicate a substantial performance improvement for the example dataset, which is unreasonable in view of the very small modification of the reference set (2 of 1,967,069 citations). There is no change for C and F. Again the relevance of the counting rule is demonstrated: chosing the L rule leads to unreasonable behavior.

It is tempting to conclude that counting rule C yields a good solution. However, it is also not immune to the problem that a small modification in the reference set can lead to unreasonable changes in the performance evaluation. To demonstrate this, I turn to the year 2005. There are 743 papers with 31 citations tied at the 90% threshold. As Table 3 shows, this threshold is approximately at the middle of the uncertainty interval. Thus a minor modification does not influence the L and E values, but can shift the average percentile value of the tied publications in the reference set from just above to just below the threshold, yielding a large change of the C value for the reference set. In the present example, it is sufficient to assume that 10 of the border line papers had each received one citation less. As a consequence, the C value for the example dataset indicates a substantially improved performance. The F result increases slightly like in Table 1.

**Concluding remarks**

In the present case study the performance of Europhysics Letters (EPL) has been analyzed for 25 years by comparison with the citation distribution of all physics papers in the Web of Science. Various counting rules have been applied. It was shown that they lead to different results depending on the attribution of the papers with the same number of citations at the borders between the PRCs. The fractional scoring approach leads to exactly 10% HCPs in the reference set. For the other counting rules I have found substantial deviations from the 10% aim in the reference set. Consequently there are substantial differences found between the results of the various counting rules for the evaluated example datasets. In specific situations unreasonable behavior was demonstrated for the L, E, and C counting rules, caused by small modifications of the reference set. This unreasonable behavior points to flaws in these counting rules. I have already pointed out elsewhere (Schreiber, 2012b) that the fractional scoring method requires a change of perspective, because papers are attributed to percentile intervals and not to discrete points on the percentage scale. Tied papers are given the same interval. These intervals are then utilized to assign weights for the evaluation of a specific dataset in comparison with the reference set. Consequently, it is not important for the evaluation, how the tied papers can be ranked. Nevertheless, if such a ranking is additionally asked for, it appears to be reasonable for all practical purposes to assign the highest possible rank among the tied papers to all tied papers. This is not only the easiest solution, but it seems to be widely accepted in daily life, e.g. in the ranking of clubs in soccer league tables.

Another problem of the fractional scoring method concerns the case of papers with zero citations. If there are many such papers tied, the corresponding uncertainty interval could be so large that it does not completely fall into the lowest PRC. Consequently, some fraction of the tied papers would be assigned a weight from one (or even more than one) higher PRC. This might be viewed as an unsatisfactory side-effect, but it cannot be avoided, if one aims at always getting the exact threshold percentages for a reference set. In the case of HCPs the zero-citation-problem is very unlikely to occur, because it would be relevant only for datasets with more than 90% uncited papers. But if one uses, e.g., 100 PRCs in the R100 indicator (Leydesdorff et al. 2011), then it is not unlikely that datasets with more 1% uncited papers occur thus exceeding the border of the lowest PRC. To be specific, if there are 5% uncited papers in such a case, these would be attributed fractionally to the five lowest PRCs and would get assigned fractionally the respective weights. Consequently, each paper would get a weight which is higher than that for the bottom PRC. This is somewhat unusual and might be considered odd but it avoids the similarily odd problem of a nonhomogeneous distribution of reference-set papers to the 100 PRCs, which occurs when one applies the standard procedure and assigns more than 1 % uncited papers to the lowest PRC. If one considers the example with 5% uncited papers as a reference set, one would thus obtain a gap with 4 empty PRCs.

A sufficiently large number of uncited papers leads to problems for the counting rule E, too. If there are more than 1% uncited papers in the reference set, then the lowest PRC of the 100 PRCs in the R100 indicator would be empty, unless one specifically demands that all uncited papers are assigned to the lowest PRC in violation of the counting rule. Similarly if there are more than 1% papers tied at the upper end of the citation distribution of the reference set, then the highest PRC in the R100 indicator would be empty when the counting rule L is applied; in this case no paper would fall into the top-1% category.

In conclusion, all the discussed methods have advantages and disadvantages. Further investigations are needed to clarify what the optimal solution to the problem of calculating percentiles and assigning papers to PRCs might be, especially for large numbers of tied papers.

**Acknowledgements**

I thank L. Waltman for his assistance in obtaining the citation data. Useful discussions with L. Bornmann, L. Leydesdorff, and L. Waltman are gratefully acknowledged.

Table 1. Percentage of HCPs and effect of a small modification (4 citations less) of the reference set for 1994.

|  | L | E | C | F |
|---|---|---|---|---|
| reference set | 9.73 | 10.00 | 10.00 | 10.00 |
| modified reference set | 10.00 | 10.33 | 10.00 | 10.00 |
| example set | 9.76 | 10.20 | 10.20 | 10.19 |
| example, mod.ref. | 10.20 | 10.64 | 10.20 | 10.20 |

Table 2. Same as Table 1, but for 2 citations less in 2003.

|  | L | E | C | F |
|---|---|---|---|---|
| reference set | 9.62 | 10.00 | 10.00 | 10.00 |
| modified reference set | 10.00 | 10.38 | 10.00 | 10.00 |
| example set | 5.45 | 6.03 | 6.03 | 6.03 |
| example, mod.ref. | 6.03 | 6.03 | 6.03 | 6.03 |

Table 3. Same as Table 1, but for 10 citations less in 2005.

|  | L | E | C | F |
|---|---|---|---|---|
| reference set | 9.72 | 10.28 | 9.72 | 10.00 |
| modified reference set | 9.72 | 10.27 | 10.27 | 10.00 |
| example set | 8.56 | 9.42 | 8.56 | 8.98 |
| example, mod.ref. | 8.56 | 9.42 | 9.42 | 8.99 |

**Fig. 1** Percentage intervals of publications with citation frequencies below (grey), at (orange and red), and above (yellow) the 90% threshold for the reference datasets; the number of citations at the threshold is indicated on the right

**Fig. 2** Same as Figure 1, but for the investigated example datasets. The black lines between the orange and red bars indicate the F results

**Fig. 3** Deviations X/F-1 (in percent) of the X = L, E, and C values from the fractional scoring results F of the HCPs indicator for the example datasets

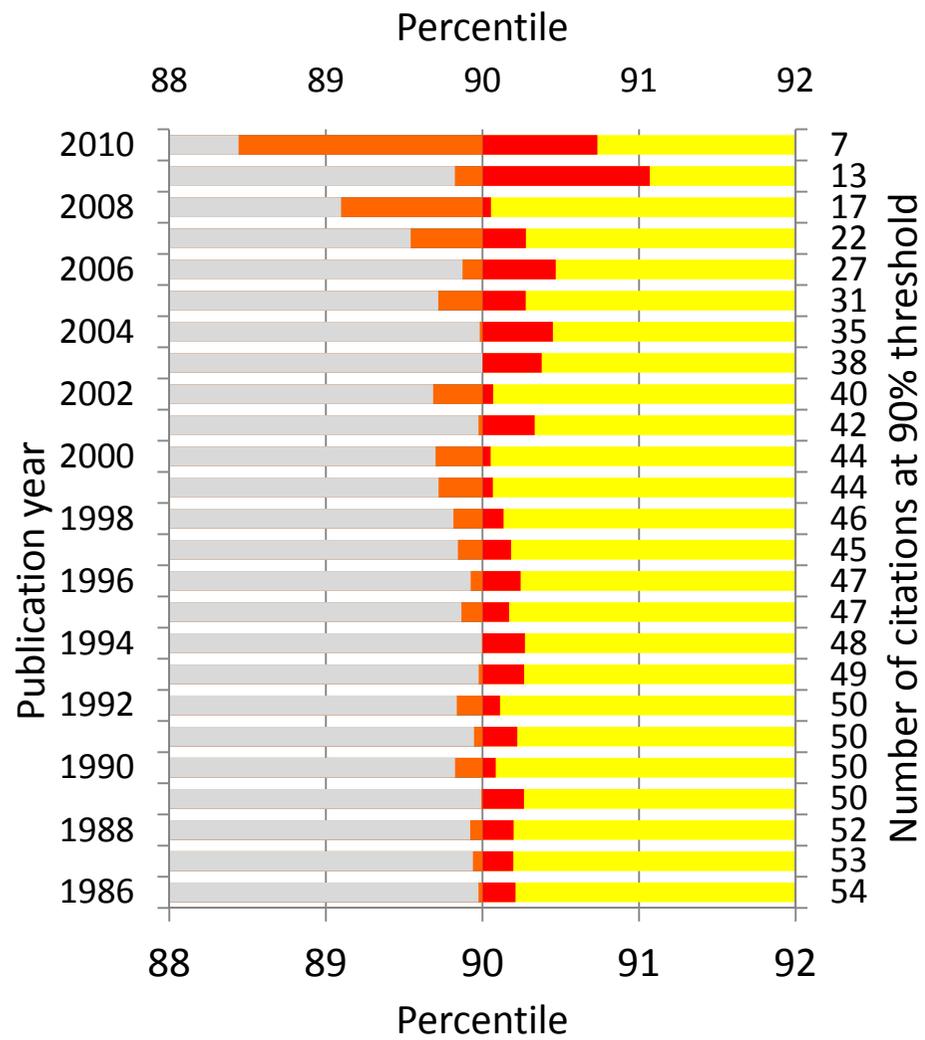

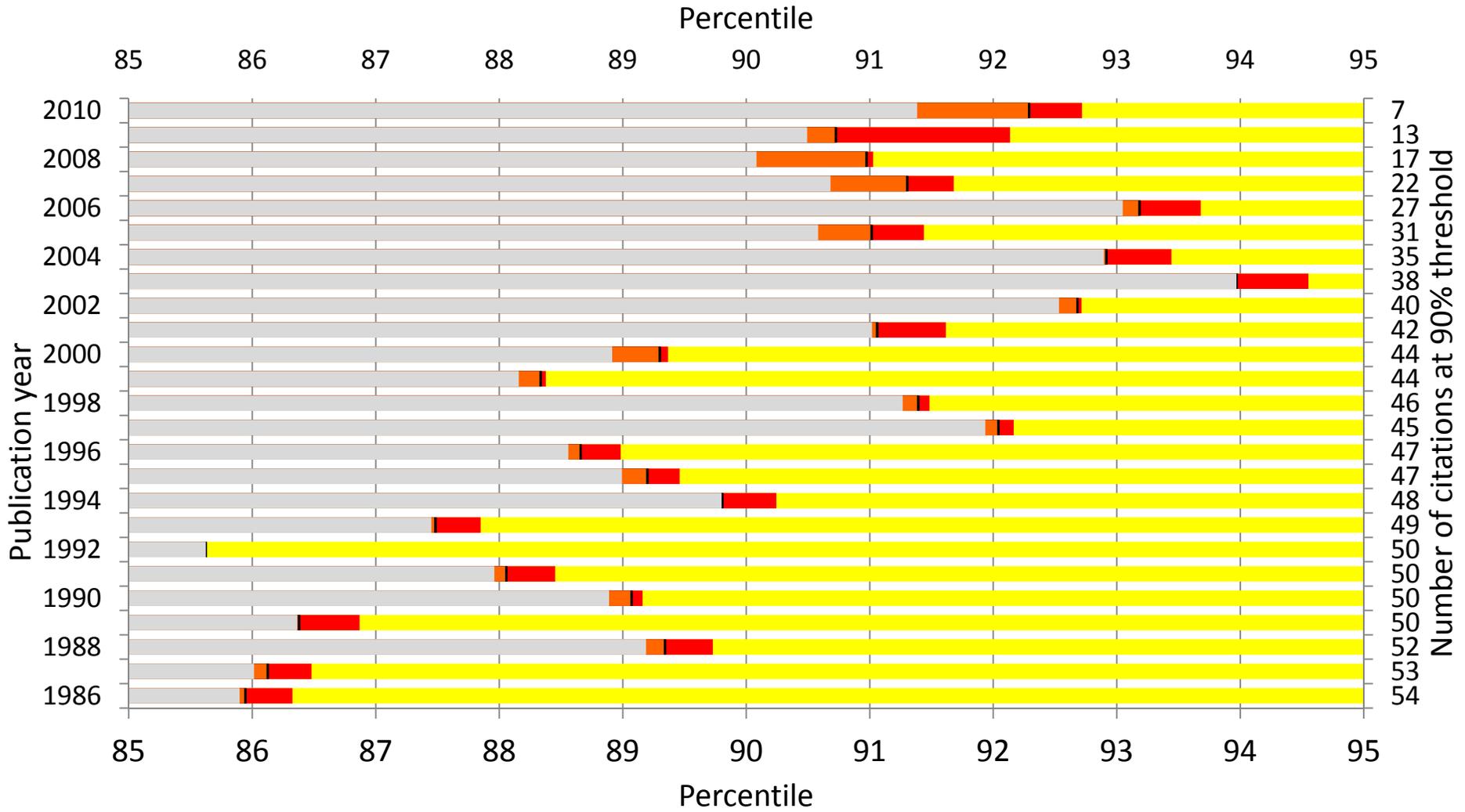

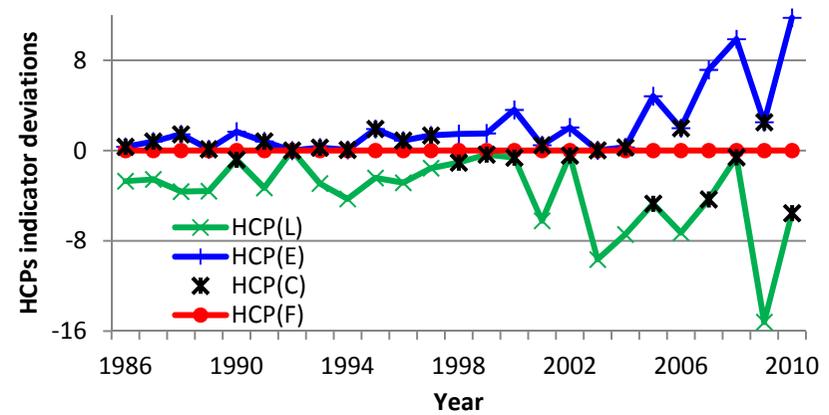